\documentclass[preprint,showpacs,preprintnumbers,amsmath,amssymb,endfloats*]{revtex4}
\usepackage{graphicx}

\begin{document}

\thispagestyle{empty}
\title{Control of the Casimir force by the modification
of dielectric properties with light
}

\author{F.~Chen${}^{1}$, G.~L.~Klimchitskaya${}^2$,
V.~M.~Mostepanenko${}^3$, and U.~Mohideen${}^1$}

\affiliation{${}^{1}$Department of Physics and Astronomy,
University of California,
Riverside, California 92521, USA. \\
${}^2$North-West Technical
University, Millionnaya St. 5, St.Petersburg, 191065, Russia.\\
${}^3$Noncommercial Partnership ``Scientific Instruments'',
Tverskaya St.{\ }11, Moscow, 103905, Russia.}

\begin{abstract}
The experimental demonstration of the modification of the Casimir 
force between a gold coated sphere and a single-crystal Si membrane 
by light pulses is performed. The specially designed and fabricated 
Si membrane  was irradiated with 514\,nm laser pulses of 5\,ms 
width in high vacuum leading to a change of the charge-carrier 
density. The difference in the Casimir force in the presence and in 
the absence of laser radiation was measured by means of an atomic 
force microscope as a function of separation at different powers of 
the  absorbed light. The total experimental error of the measured 
force differences at a separation of 100\,nm varies from 10 to 20\% 
in different measurements. The experimental results are compared 
with theoretical computations using the Lifshitz theory at both 
zero and laboratory temperatures. The total theoretical error 
determined mostly by the uncertainty in the concentration of charge 
carriers when the light is incident is found to be about 14\% at 
separations less than 140\,nm.  The experimental data are consistent 
with the Lifshitz theory at laboratory temperature, if the static
dielectric permittivity of high-resistivity Si in the absence of 
light is assumed to be finite. If the dc conductivity of 
high-resistivity Si in the absence of light is included into the 
model of dielectric response, the Lifshitz theory at nonzero 
temperature is shown to be experimentally inconsistent at 95\% 
confidence. The demonstrated phenomenon of the modification of
the Casimir force through a change of the charge-carrier density 
is topical for applications of the Lifshitz theory to real 
materials in fields ranging from nanotechnology and condensed 
matter physics to the theory of fundamental interactions.
\end{abstract}


\maketitle

\section{Introduction}

After many years of pure academic research, the Casimir effect
\cite{1} is presently of much interest in connection with
applications in nanomechanical devices \cite{2,3,4}, noncontact
friction \cite{4a,5,6,7},
carbon nanotubes \cite{B1,B2,B3,B4},
Bose-Einstein condensation \cite{8,9}
and for constraining predictions of modern unification theories
of fundamental interactions \cite{10,11,12,13,13a}. These areas of
application were made possible by extensive experimental
investigation of the Casimir force \cite{12,13,13a,14,15,16,17,18,19,20}
and the generalization to real materials of field-theoretical methods
which were applicable to only idealized boundaries (see reviews \cite{21,22}).

The basic theory of the Casimir and van der Waals forces at nonzero
temperature proposed by Lifshitz \cite{23,24} allows one to calculate
all quantities of physical interest using the dielectric
permittivity of boundary materials along the imaginary frequency axis.
This theory was originally developed for the configuration of two
semispaces and was later extended for any layer structure
\cite{25,26,27}. Using the proximity force theorem \cite{28},
Lifshitz-type formulas for the configuration of a sphere or
a cylinder above a plate were obtained and successfully used for
the interpretation of experimental data \cite{3,12,13,13a,14,15,16,17,18,20}.
For a long time, the lack of exact results for these configurations
made it possible to question the validity of the comparison of
experiment and theory based on the proximity force theorem.
Recently, however, both the exact analytical \cite{29,30,31} and
numerical \cite{32} results for the Casimir force between a sphere
(cylinder) and a plate were obtained demonstrating that at small $z$
the corrections to the proximity force theorem for both
configurations are in fact less than $z/R$ ($z$ is the
separation between a cylinder or a sphere of radius $R$ and a plate),
i.e., less than it was supposed in the comparison of experiment with
theory. Thus, the use of the proximity force theorem in
Refs.~\cite{3,12,13,13a,14,15,16,17,18,20} and below is substantiated on the
basis of first principles of quantum field theory.

The vital issue in many applications of the Casimir effect is how
to control the magnitude of the force by changing the parameters
of the system. In this respect the possibility that the Casimir
force can change sign from attraction to repulsion depending on
system geometry is of much importance. It may be used to prevent
collapse of small mechanical elements onto nearby surfaces
in nanodevices \cite{33}.
However, the Casimir repulsion has yet to be observed
experimentally. An alternative method to control the magnitude of the
Casimir force is to change the material properties of the
interacting bodies. In Ref.~\cite{34} the Casimir force was
measured acting between a plate and a sphere coated with a
hydrogen-switchable mirror that become transparent upon hydrogenation.
Despite expectations, no significant decrease of the Casimir force
owing to the increased transparancy of the plates was observed.
The negative result is explained by the Lifshitz theory which
requires the change of the reflectivity properties within a wide
range of frequencies in order to markedly affect the magnitude of
the Casimir force. This requirement is not satisfied by the
hydrogenation.

All modern experiments on the measurement of the Casimir force
mentioned above \cite{3,12,13,13a,14,15,16,17,18,19,20,34} used
metallic test bodies. Metallic surfaces are necessary to reduce
and compensate the effects of residual charges and work function
differences.
 It is, however, hard to modify their reflectivity
properties over a sufficiently wide range of frequencies.
The appropriate materials for the control, modification and fine
tunning of the Casimir force are semiconductors. The reflectivity
properties of semiconductor surfaces can be changed in a wide
frequency range by changing the carrier density through the
variation of temperature, using different kinds of doping or,
alternatively, via the illumination of the surface with laser light.
At the same time, semiconductor surfaces with reasonably high
conductivity avoid accumulation of excess
charges and, thus, preserve the advantage of
metals. In addition as  semiconductors are the basic fabrication
materials for nanotechnology, the use of semiconductor surfaces
for the control of the Casimir force will lead to many applications.

The modification of the Casimir force between a gold coated plate
and sphere, attached to the cantilever of an atomic force
microscope (AFM), through the variation of temperature was
considered in Ref.~\cite{35}. While changing the temperature to
modify the carrier density in semiconductors is a good method
in theory, it leads \cite{36} to large systematic errors in the
measurement setup using the AFM. In Ref.~\cite{37} the Casimir
force between a gold coated sphere and a single crystal B-doped
Si plate was measured in high vacuum. It was found that the force
between a metal and a semiconductor decreases with increase of
separation more quickly than between two metals. In Ref.~\cite{38}
the experimental data for the Casimir
force between a gold coated sphere and B-doped Si plate were
compared with two different theoretical computations, one made
for the B-doped Si used and another one for high-resistivity Si.
It was shown that the computation using the tabulated optical data
for high-resistivity Si is excluded by experiment at 70\%
confidence while the theoretical results computed for the plate
used in experiment are consistent with data.
In Ref.~\cite{380} the difference in the Casimir forces
between a gold coated sphere and two P-doped
Si plates with different charge-carrier densities was
directly measured at a 95\% confidence level.
This demonstrates
that the change of carrier density due to doping leads to
noticeable modification of the Casimir force.

The most suitable method to change the carrier density in
semiconductors is through the illumination of the surfaces by laser light
(see, e.g., \cite{38a,38b}). An early attempt to measure the
van der Waals and the
Casimir forces between semiconductors and modify them with light
was reported in Ref.~\cite{39}. Attractive forces were measured
between a glass lens and a Si plate and also between the glass
lens coated with amorphous Si and the Si plate. However, the
glass lens is an insulator and therefore the electric forces such
as due to work function potential differences could not be
controlled. This might also explain that no force change
occured on illumination at separations below 350\,nm \cite{39}
where it should have been most pronounced.

The present paper contains the detailed results of our experiments
on the modification of the Casimir force by the irradiation of
a Si membrane with laser pulses. The first observation of this effect
at only one absorbed power
was briefly reported in Ref.~\cite{40}. Here we report new
measurements performed at different absorbed powers and
the theoretical analysis on the accuracy of the obtained results
and on the comparison of experiment
with theory. In our experiments the carrier density in the Si
membrane is changed by the incident light, and the difference in the
Casimir force acting between that membrane and the gold coated
sphere in the presence and in the absence of light is measured.
The experimental error of difference force measurements
for the different absorbed powers determined at a 95\% confidence level
varies between 10 to 20\% at a separation of 100\,nm and increases
with the increase of separation. The measurement data collected
at different powers of the incident light at the laboratory
temperature $T=300\,$K were compared with the Lifshitz theory
at both zero and at laboratory temperatures. The data are shown to be
consistent with theory at laboratory temperature if in the
absence of light the static dielectric permittivity of Si is
assumed to be finite. The Lifshitz theory
at laboratory temperature taking account of the
dc conductivity of high-resistivity Si in the absence of light is
excluded experimentally at a 95\% confidence level. Thus, our experiments
not only demonstrate the modification of the Casimir force through
the irradiation of a semiconductor surface, but also lead to the
important result that the inclusion of zero-frequency  conductivity
of high-resistivity Si  in the model
of dielectric response results in a contradiction
between the Lifshitz theory at laboratory temperature
and experiment.
This contradiction is caused by different contributions from the
reflection of the transverse magnetic mode on a Si surface at zero frequency
in the absence and in the presence of conductivity.
The obtained conclusion supports recent theoretical
results that the inclusion of dielectric dc conductivity for the
dielectric-dielectric \cite{7} and
 dielectric-metal \cite{IJMPA,48a} configurations at nonzero temperature leads
to contradiction between the Lifshitz theory and the Nernst heat
theorem, and thus such inclusion is impermissible.
At the same time, the experimental data are shown to be
consistent with the Lifshitz theory at zero temperature,
irrespective of whether the dc conductivity of high-resistivity
Si is included or not.

The paper is organized as follows. In Sec.~II the experimental
setup and sample preparation are described. Sec.~III contains the
description of the measurement procedure and obtained experimental
results. This includes the calibration of the setup, the measurement
of the lifetime of excited carriers, the measurement of the
difference in the Casimir force when the light is on and off, and
the analysis of the experimental errors. In Sec.~IV the experimental
results are compared with the theory. Here the difference in the Casimir
force with and without incident laser light is calculated and the
theoretical
errors are analyzed. By combining the experimental and theoretical
errors, the quantitative measure of agreement between experiment and
theory at 95\% confidence is presented. Sec.~V analyzes the role of
the dc conductivity of high-resistivity Si in the Casimir
force. Sec.~VI contains our conclusions and discussion.

\section{Experimental setup and sample preparation}

Here we discuss the experimental setup used to demonstrate the
modification of the Casimir force through the radiation induced change
in the carrier density. The general scheme of the setup is shown in
Fig.~\ref{setup}. A high vacuum based AFM was employed to measure
the change in the Casimir force between a gold coated sphere of
diameter $2R=197.8\pm 0.3\,\mu$m and a Si membrane (coloured black)
in the presence and in the absence of incident light. An oil free vacuum
chamber with a pressure of around $2\times 10^{-7}\,$Torr was used.
The polystyrene sphere coated with a gold layer of $82\pm 2\,$nm thickness
was mounted at the tip of a 320$\,\mu$m conductive cantilever
(see Fig.~\ref{setup}). The Si membrane (see below for the process of its
preparation) was mounted on top of a piezo which is used to change the
separation distance $z$ between the sphere and the membrane from
contact to 6$\,\mu$m. The excitation of the carriers in the
Si membrane was
done with 5\,ms wide light pulses (50\% duty cycle). These pulses were
obtained from a CW Ar ion laser light at 514\,nm wavelength  modulated
at a frequency of 100\,Hz using an Acousto-Optic-Modulator (AOM).
The AOM is triggered with a function generator. The laser pulses were
focused on the bottom surface of the Si membrane. The Gaussian width of
the focused beam on the membrane was measured to be $0.23\pm 0.01$mm.

The cantilever of the AFM flexes when the Casimir force between the sphere
and the membrane changes depending on the presence or the absence of
incident light on the membrane. This cantilever deflection is monitored with
a 640\,nm beam from an additional laser (see Fig.~\ref{setup})
reflected off the top of the cantilever tip. An optical filter was used
to prevent the interference of the 514\,nm excitation light with the
cantilever deflection signal.
The transmission of this filter at 514\,nm is
0.001\%. Including the less than
1\% transmission through the
Si membrane and the diode solid angle
of $10^{-4}$, the impact of the 514\,nm light
leakage leads to less than $10^{-6}$\,pN changes in the
force difference. These changes are negligibly small as
compared with the measured cantilever deflection signal.
The latter leads to a difference signal
between the two photodiodes. The resulting modification of the Casimir
force in response to the carrier excitation is measured with a lock-in
amplifier. The same function generator signal used to generate the Ar laser
pulses is also used as a reference for the lock-in amplifier.

The most important part
of the setup is the Si membrane. It should be sufficiently thin and of
high resistivity to ensure that the density of charge carriers increases
by several orders of magnitude under the influence of the laser pulses.
The Si membrane should be thick enough to make negligible the photon
pressure of the transmitted light, as the illumination is incident
on the bottom surface of the membrane (see Sec.~IVB for details).
The thickness of the
Si membrane has to be greater than 1\,$\mu$m, i.e., greater than the
optical absorption depth of Si at the wavelength of the laser pulses.
Fabrication of the few micrometer thick Si membrane with the necessary
properties is described below.

A commercial Si grown on insulator wafer (SOI) was used
as the initial product. The insulator in this case is SiO${}_2$ which
is the native oxide of Si and thus leads to only small reductions of
the excited carrier lifetime in Si.
A layout of the wafer is shown in Fig.~2.
The wafer consists of a Si substrate of 600$\,\mu$m thickness
and a Si top layer of 5$\,\mu$m thickness (both are single crystals
and have a $\langle 100\rangle$ crystal orientation) with the
buried
intermediate SiO${}_2$ layer of thickness 400\,nm
(see Fig.~2a). The Si is
p-type doped with relatively high nominal resistivity of about
10$\,\Omega\,$cm. The corresponding carrier density is equal to
$\tilde{n}\approx 5\times 10^{14}\,\mbox{cm}^{-3}$ \cite{41}.

The thickness
of the Si substrate is reduced to about 200$\,\mu$m through mechanical
polishing. Then, after RCA cleaning of the surface, the wafer is oxidized
at $T=1373\,$K in a dry O${}_2$ atmosphere for a duration of
72 hours. As a result, in addition to the buried SiO${}_2$ layer, a
 thermal oxide layer with a thickness of about 1$\,\mu$m is
formed on both (bottom and top) sides of the wafer (Fig.~2b).
This oxide layer serves as
a mask for subsequent Tetra Methyl Ammonium Hydroxide (TMAH) etching of
the Si. First, a hole with the diameter of 0.85\,mm is etched
with HF in the center of the
bottom oxidation layer (Fig.~2c). This exposes the Si substrate.
Next, TMAH is used at 363\,K to etch the Si substrate through the hole
formed in the oxide mask (Fig.~2d). Note that TMAH selectively etches
Si as its etching rate for Si is 1000 times greater than
for SiO${}_2$. TMAH etching leads to the formation of a hole through the
Si substrate. Given the selectivity of the etching,
the buried 400\,nm oxide is the stop etch layer.
Finally, all the thermal oxidation layers and buried oxidation layer in the
hole are etched away in HF solution to form a clean Si membrane over
the hole as in Fig.~2e. The thickness of this membrane
was measured to be  $4.0\pm 0.3\,\mu$m
using an optical microscope. In order for voltages to be applied to
the Si membrane,
an ohmic contact is formed by a thin film of Au deposited on the edge of
the membrane followed by annealing at 673\,K for 10 min. The Si membrane
was cleaned with Nanostrip and then passivated by
dipping in 49\% HF for 10\,s. The passivated Si membrane was then
mounted on top of the piezo as described above.

\section{Measuring procedure and experimental results}

\subsection{Calibration of the setup}

All calibrations and other measurements are done at the same period of
time as the measurement of the difference of Casimir forces and in
the same high vacuum apparatus. The calibration of the deflection
signal of the cantilever from the photodiodes, $S_{\rm def}$, and
the determination of the separation on contact and residual potential
difference between the gold coated sphere and Si membrane is done by
measuring the distance dependence of an applied electrostatic force.
For this purpose the same function generator (see Fig.~\ref{setup})
is used for applying voltages to the membrane. For an attractive force
$S_{\rm def}<0$ and can be measured either as a current or a voltage.
In addition, a small correction has to be applied to the separation
distance between the gold sphere and the Si membrane due to the
movement of the cantilever. The actual separation distance $z$
between the bottom of the sphere and the membrane is given by
\begin{equation}
z=z_{\rm piezo}+mS_{\rm def}+z_0.
\label{eq1}
\end{equation}
\noindent
Here $z_{\rm piezo}$ is the distance moved by the piezo, $m$ is
the deflection coefficient in units of nm per unit deflection
signal, and $z_0$ is the average separation on contact of the
gold surface and Si membrane. $z_0$ is nonzero due to the stochastic
roughness of the surfaces. The complete movement of the piezo was
calibrated using a fiber optic interferometer. To extend and
contract the piezo, continuous triangular voltages between
0.01--0.02\,Hz are applied to it. Given that the experiment is
done at room temperature, applying of static voltages would lead
to piezo creep and loss of position sensitivity. The deflection
coefficient $m$ can also be measured by the application of
electrostatic forces between the sphere and the membrane.

In our measurements, the gold sphere was kept grounded. The electric
contact to the sphere was accomplished by applying a very thin gold
coating to the cantilever. The electrostatic force between the sphere
and the membrane is given by \cite{42}
\begin{equation}
F_e(z)=2\pi\varepsilon_0(V-V_0)^2
\sum\limits_{n=1}^{\infty}
\frac{\coth\alpha-n\coth n\alpha}{\sinh n\alpha},
\label{eq2}
\end{equation}
\noindent
where $V$ is the voltage applied to the
Si membrane, $V_0$ is the residual potential difference between the
grounded sphere and membrane, $\cosh\alpha=1+z/R$,
and $\varepsilon_0$ is the permittivity of vacuum.
The nonzero value of $z$ at contact, $z_0$, is due to the
surface roughness. In the complete
 measurement range of the electrostatic
force from contact to 6\,$\mu$m,
Eq.~(\ref{eq2}) can be rearranged to the following more simple
form within the limits of relative error less than $10^{-4}$
\cite{38}:
\begin{equation}
F(z)=-2\pi\varepsilon_0(V-V_0)^2
\sum\limits_{i=-1}^{6}
c_i
\left(\frac{z}{R}\right)^i\equiv X(z)(V-V_0)^2,
\label{eq3}
\end{equation}
\noindent
where
\begin{eqnarray}
&&
c_{-1}=0.5,\quad c_0=-1.18260,\quad c_1=22.2375,\quad
c_2=-571.366,
\nonumber \\
&&
c_3=9592.45,\quad c_4=-90200.5,\quad c_5=383084,
\quad c_6=-300357.
\nonumber
\end{eqnarray}

First, 30 different dc voltages between 0.65 to --0.91\,V are applied
to the Si membrane. The cantilever deflection signal is measured as a
function of the distance. The 0.02\,Hz triangular wave was applied
to the piezo to change the distance between the sphere and the membrane
over a range of 6\,$\mu$m. Larger applied voltages lead to more
cantilever deflection and, according to Eq.~(\ref{eq1}), to a
contact of the two surfaces at larger $z_{\rm piezo}$. The dependence
of $z_{\rm piezo}$ at contact of the sphere and the membrane on the
applied voltage can then be used to measure the deflection
coefficient $m$. In order to determine the contact of the two
surfaces precisely, 32768 data points at equal time intervals were
acquired for each force measurement (i.e., the interval between two
points was about 0.18\,nm). In cases, where the contact point was
between two neighboring data points, a linear interpolation was used
to identify the exact value. The deflection coefficient was found
to be $m=137.2\pm 0.6\,$nm per unit deflection signal. The difference
in the value of $m$ from previous measurements \cite{18,37,38} is
due to the use of the 514\.nm filter which reduced the cantilever
deflection signal. The obtained value of $m$ was used to
correct the separation distance in all measurements in accordance
with Eq.~(\ref{eq1}). The electrostatic force resulting from the
application of the dc voltages is also used in the determination
of the separation on contact of the two surfaces. The fit of the
experimental force-distance relation to the theoretical Eq.~(\ref{eq3})
is done as outlined in our previous work \cite{18,37,38}.
The separation distance on contact was determined to be $z_0=97\,$nm.
The uncertainty in the quantity $z_0+mS_{\rm def}$ from Eq.~(\ref{eq1})
was found to be 1\,nm. This leads to the same error in absolute
separations $\Delta z=1\,$nm because the error in piezo calibration
is negligibly small.

For the calibration of the deflection signal and the determination of
the residual potential difference between the two surfaces, an
improved method, rather than simple application of the dc voltages
to the membrane was used. This was done to avoid systematic errors due
to scattered laser light. In addition to the application of the dc
voltage to the membrane, described above, square voltage pulse of
amplitudes from 1.2 to --0.6\,V and time interval corresponding to
a separation distance between 1 to 5\,$\mu$m was also applied to
the membrane. Fig.~\ref{deflSignal} shows the deflection signal of
the cantilever in response to both the applied dc voltage and the square
pulse as a function of the separation distance between the gold sphere
and Si membrane. By measuring only the difference in signal during the
pulse allows one to avoid the need for a background subtraction.
The fit of the difference signal to Eq.~(\ref{eq3}) leads to the value
of the signal calibration constant $6.16\pm 0.04\,$nN per unit
deflection signal. The same fit was used to determine the residual
potential difference between the sphere and the membrane which was found
to be $V_0=-0.171\pm 0.002\,$V. The large width of the pulse applied
in addition to the dc voltage allowed confirmation of the distance
independence of the obtained values of the calibration constant and
the residual potential difference.

\subsection{Excited carrier lifetime measurement}

An independent measurement of the lifetime of the carriers excited in the
Si membrane by the pulses from the Ar laser was performed. For this purpose
a non-invasive optical pump-probe technique was used \cite{43,44}.
The same Si membrane and Ar laser beam modulated by the AOM at 100\,Hz
to produce 5\,ms wide square light pulses, as used in the Casimir force
measurement, were employed as the sample and the pump, respectively.
The diameter of
the pump beam on the sample was measured to be $0.72\pm 0.02\,$mm.
A CW beam with a 1\,mW power at a wavelength of 1300\,nm was used
as a probe. The probe beam photon energy is below the band gap energy
of Si and is thus not involved in carrier generation. This beam was
focused to a Gaussian width size $w_0=0.135\pm 0.003\,$mm. Thus the
focal spot size of the probe beam is much smaller than the focal spot
size of the pump light. This allowed one to measure the lifetime in
a homogeneous region of excited carriers. The change in the reflected
intensity of the probe beam in the presence and in the absence of
Ar laser
pulse was detected with a InGaAs photodiode. The change
in reflected power of
the probe beam was monitored as a function of time in an oscilloscope
and found to be consistent with the change of carrier density.
Near normal incidence for the pump and probe beams was used, with
care taken to make sure that the InGaAs photodiode was isolated from
the pump beam. The time decay of the reflected probe beam in response
to the square Ar light pulses is shown in Fig.~\ref{lifetime}.
The change of the reflectivity of the probe is fit to an exponential
of the form $-\exp(-t/\tau)$ where $\tau$ is the effective carrier
lifetime. By fitting the whole 5\,ms decay of the change in reflected
power, the effective excited carrier lifetime was measured to be
$\tau=0.47\pm0.01\,$ms. Note that this time represents both surface
and bulk recombination and is consistent with that expected for Si.
Some dependence of the lifetime of the excited carriers on their
concentration was observed. In the first 0.5\,ms, while the
concentration is still high enough, the average value of the
excited carrier lifetime was measured to be $\tau=0.38\pm 0.03\,$ms.
The measured values of the carrier lifetime will be used in
Sec.~IVA in the theoretical computations of the Casimir force
differences for the comparison with several measurements
having varying power of Ar laser.

\subsection{Experimental results and error analysis}

Here we present the determination of the difference in the Casimir
force resulting from the irradiation of the Si membrane with
514\,nm laser pulses.
In fact it is the difference in the total force (Casimir and
electric) which is measured. As was indicated above, even with no
applied voltages there is some residual potential difference $V_0$
between the sphere and the membrane. The preliminary value of $V_0$
was determined during the calibration of the setup in the absence
of laser pulses. In the presence of pulses (even during the dark phases
of a pulse train) the values of the residual potential difference can
be different. We represent these residual potential differences
during the bright and dark phases of a laser pulse train
(the latter is not exactly equal to the one determined in calibration)
$V_0^l$ and $V_0$, respectively. During the bright phases
of the pulse train we apply to the Si membrane the voltage $V^l$ and
during the dark phases the voltage $V$. Using Eq.~(\ref{eq3}) for the
electric force, we can represent the difference in the total force
(electric and Casimir) for the states with and without
carrier excitation in the following form:
\begin{equation}
\Delta F_{\rm tot}(z)=X(z)\left[(V^l-V_0^l)^2-(V-V_0)^2\right]+
\Delta F_C(z).
\label{eq4}
\end{equation}
\noindent
Here
\begin{equation}
\Delta F_C(z)=F_C^l(z)-F_C(z)
\label{eq5}
\end{equation}
\noindent
is the difference in the Casimir force and $F_C^l(F_C)$ is the
Casimir force with (without) light. The difference in the total
force in Eq.~(\ref{eq4}) was measured by the lock-in amplifier with
an integration time constant of 100\,ms which corresponds to a
bandwidth of 0.78\,Hz. The measurement procedure is
described below.

First we kept $V=\mbox{const}$ and changed $V^l$. The parabolic
dependence of $\Delta F_{\rm tot}$ on $V^l$ in Eq.~(\ref{eq4})
was measured at different separations $z$. Care should be taken to
apply only small voltage amplitudes (up to a few tens of mV) so as
to keep the space charge region negligible. At every measured
separation distance  $\Delta F_{\rm tot}$ is plotted as a function
of $V^l$. As is seen from Eq.~(\ref{eq4}), the value of $V^l$
where the parabola reaches a maximum is $V_0^l$ [recall that
$X(z)<0$]. In this way the value $V_0^l=-0.303\pm 0.002\,$V was
found and shown to be independent of the separation
from 100 to 500\,nm where the difference in the Casimir force can
be measured. Next we kept $V^l=\mbox{const}$, changed $V$
and measured the parabolic dependence of $\Delta F_{\rm tot}$ on $V$
at different separations. The value of $V$ where parabolas reach minima
is $V_0=-0.225\pm 0.002\,$V. These values of the residual
potential difference between the sphere and the membrane in the presence
and in the absence of excitation light were substituted in
Eq.~(\ref{eq4}).
The small change of around 78\,mV
in the residual potential difference
between the sphere and the membrane in the
presence and in the absence of excitation light
is primarily due to the screening of surface
states by few of the optically excited electrons and
holes. The above small value is equal to the change in band
bending at the surface. It is consistent with
the fact that almost flat bands are obtained at the
surface with the surface passivation technique used
(see, e.g., \cite{R1,R2}).

Then other voltages ($V^l,V$) were applied to the Si membrane and the
difference in the total force $\Delta F_{\rm tot}$ was measured as
a function of separation. Data were collected from contact at equal
time intervals corresponding to 3 points per 1nm (i.e., in 1209 points
within the separation interval from 100 to 500\,nm). From these
measurement results, the difference in the Casimir force
$\Delta F_C^{\rm expt}(z)$ was determined from Eq.~(\ref{eq4}).
This procedure was repeated with some number of pairs ($J$) of
different applied voltages ($V^l,V$) and at each separation the
mean value $\langle\Delta F_C^{\rm expt}(z)\rangle$ was found.
In Fig.~\ref{result1} the experimental data for
$\langle\Delta F_C^{\rm expt}(z)\rangle$ as a function of separation are shown
by dots for different absorbed laser powers: $P^{\rm eff}=9.3\,$mW
($J=31$), 8.5\,mW ($J=41$), 4.7\,mW ($J=33$) in figures a, b, and
c, respectively.
The corresponding incident powers were 15.0, 13.7 and 7.6\,mW, respectively.
As expected, the magnitude of the Casimir force
difference has the largest values at the shortest separations and
decreases with the increase of separation. It also decreases with the
decrease of the absorbed laser powers (the solid, short- and long-dashed
lines in Fig.~\ref{result1} are explained in Sec.~IV devoted to the
comparison with theory).

Now we proceed with the analysis of the experimental errors.
The variance of the mean difference in the Casimir force is
defined as
\begin{equation}
s_{\langle\Delta F_C^{\rm expt}\rangle}(z_i)=\left\{
\frac{1}{J(J-1)}\sum\limits_{j=1}^{J}
\left[\Delta F_C^{\rm expt}(z_{ij})-
\langle\Delta F_C^{\rm expt}(z_i)\rangle
\right]^2\right\}^{1/2},
\label{eq6}
\end{equation}
\noindent
where $i$ is the number of point in one set of measurements changing
from 1 to 1209, $j$ is the number of the pair of the applied voltages.
Using Student's $t$-distribution with a number of degrees of freedom
$f=30$ (or 40 and 32 for the measurements with different absorbed powers)
and choosing $\beta=0.95$ confidence, we obtain
$p=(1+\beta)/2=0.975$ and $t_p(f)=2.00$. Thus, the absolute random
error in the measurement of the difference Casimir force is given by
\begin{equation}
\Delta^{\rm \! rand}\left(\Delta F_C^{\rm expt}(z)\right)=
s_{\langle\Delta F_C^{\rm expt}\rangle}(z)\,t_p(f).
\label{eq7}
\end{equation}
\noindent
In this experiment the random error is separation dependent. It is
presented in Fig.~\ref{random} as a function of separation for the three
different measurements with different absorbed laser powers (lines
a, b, and c correspond to decreasing power indicated above).
As is seen from Fig.~\ref{random}, the random error is rather different
for different measurements. It is the lowest for measurement (b) which was
done with 8.5\,mW absorbed power. In this measurement the random error
decreases from 0.32\,pN at $z=100\,$nm to 0.23\,pN at $z=250\,$nm and
preserves the latter value at larger separations.

The main systematic error is due to the instrumental noise and is equal
to $\Delta_1^{\!\rm syst}(\Delta F_C^{\rm expt})\approx  0.08\,$pN
independent of separation. The systematic error
determined from the resolution error in data acquisition,
$\Delta_2^{\!\rm syst}(\Delta F_C^{\rm expt})\approx  0.02\,$pN,
also does not depend on separation. The calibration error,
$\Delta_3^{\!\rm syst}(\Delta F_C^{\rm expt})$,
depends on separation and is equal to 0.6\% of the measured difference
in the Casimir force.
These systematic errors are random quantities characterized by a
uniform distribution.
They can be combined at a given confidence probability $\beta$
with the help of statistical criterion \cite{45}
\begin{equation}
\Delta^{\!\rm syst}(\Delta F_C^{\rm expt})=\min\left\{
\sum\limits_{i=1}^{q}\Delta_i^{\!\rm syst}(\Delta F_C^{\rm expt}),
k_{\beta}^{(q)}\sqrt{
\sum\limits_{i=1}^{q}\left[
\Delta_i^{\!\rm syst}(\Delta F_C^{\rm expt})\right]^2}\right\},
\label{eq7a}
\end{equation}
\noindent
where $k_{\beta}^{(q)}$ is a tabulated coefficient.
In our experiment there are $q=3$ systematic errors listed above and
at $\beta=0.95$ (95\% confidence level)
$k_{0.95}^{(3)}=1.12$. As a result, from Eq.~(\ref{eq7a})
we arrive at the
total systematic error for all three measurements varying from
0.092 to 0.095\,pN.

The total experimental error of the force difference,
$\Delta^{\!\rm tot}\left(\Delta F_C^{\rm expt}(z)\right)$, at
95\% confidence can be found by the combination of random and
systematic errors. This is done using the statistical rule described
in Ref.~\cite{45} and applied to the Casimir force measurements in
Refs.~\cite{13,38,45a}. According to this rule, the total error is
equal to the random one if, as is the case in our experiments,
the inequality
\begin{equation}
r(z) \equiv
\frac{\Delta^{\!\rm syst}\left(\Delta F_C^{\rm expt}
(z)\right)}{s_{\langle F_C^{\rm expt}\rangle}(z)}\leq 0.8
\label{eq7b}
\end{equation}
\noindent
is satisfied.
 Thus, the total experimental error in the values of
$\Delta F_C^{\rm expt}(z)$
for all three measurements as a function
of the separation is presented in Fig.~\ref{random}.
As a result, the relative experimental error changes from 10 to 20\%
at a separation $z=100\,$nm and from 25 to 33\% at a separation
$z=180\,$nm for different absorbed laser powers. This allows us to
conclude that the modulation of the dispersion force with light is
demonstrated at a high reliability and confidence. The observed effect
cannot be due to the mechanical motion of the membrane. This is because
membrane movement due to heating (in our case less than $1{}^{\circ}$C) would
lead to a different force-distance relationship for both electrostatic
force and the Casimir force in disagreement with
our observation and the confirmation of the distance independence
of $V_0$ and $V_0^l$.
The temperature rise of less than $1^{\circ}$C is estimated based on
the net thermal energy increase in the Si
membrane.  The absorption of photons
during the course of the optical pulse increases
the thermal energy of the membrane,
while conductive and radiative heat outflow to the
Si around the membrane and surrounding leads to
a decrease in its thermal energy. The net
change results in the less than $1^{\circ}$C. The latter
would lead to a negligible
less than $10^{-6}$ relative expansion in
the diameter of the membrane.

In order to account for
roughness, the surface topography of the sphere and membrane was
characterized using the AFM. Images
resulting from the surface scan of the gold coating on the sphere
demonstrate stochastically distributed roughness peaks with heights up
to 32\,nm. Table~I contains the fractions $v_k$ of the gold coating
with heights $h_k$ ($k=1,2,\ldots 33$). The surface scan of Si surface
demonstrates much smoother relief with maximum heights equal to
1.68\,nm. The fractions $v_l$ of the Si surface with heights $h_l$
($l=1,2,\ldots,17$) are presented in Table~II.
The roughness data are used in Sec.~IV in theoretical
computations.

\section{Comparison of the experimental results with the theory}

\subsection{Calculation of the Casimir force difference}

The Casimir force acting between a large gold sphere of radius $R$ and
a plane Si membrane can be calculated by means of the Lifshitz formula
\cite{23,24,45b}, along with the use of the proximity force theorem
\cite{28,29,30,31,32}
\begin{eqnarray}
&&F_C(z)=k_BTR\sum\limits_{l=0}^{\infty}
\left(1-\frac{1}{2}\delta_{l0}\right)
\int_{0}^{\infty}k_{\bot}dk_{\bot}
\left\{\ln\left[1-r_{\|}^{(1)}(\xi_l,k_{\bot})
r_{\|}^{(2)}(\xi_l,k_{\bot})e^{-2q_lz}\right]\right.
\nonumber \\
&&\phantom{aaaaaa}\left.
+\ln\left[1-r_{\bot}^{(1)}(\xi_l,k_{\bot})
r_{\bot}^{(2)}(\xi_l,k_{\bot})e^{-2q_lz}\right]\right\}.
\label{eq8}
\end{eqnarray}
\noindent
Here $k_B$ is the Boltzmann constant. The reflectivity coefficients for
gold ($k=1$) and Si ($k=2$) for the two independent polarizations of
electromagnetic field (transverse magnetic and transverse electric modes)
are defined by
\begin{equation}
r_{\|}^{(k)}(\xi_l,k_{\bot})=
\frac{\varepsilon_l^{(k)}q_l-k_l^{(k)}}{\varepsilon_l^{(k)}q_l+k_l^{(k)}},
\quad
r_{\bot}^{(k)}(\xi_l,k_{\bot})=
\frac{k_l^{(k)}-q_l}{k_l^{(k)}+q_l},
\label{eq9}
\end{equation}
\noindent
where $\xi_l=2\pi k_BTl/\hbar$ are the Matsubara frequencies,
$\varepsilon_l^{(k)}=\varepsilon^{(k)}(i\xi_l)$,
$\varepsilon^{(k)}(\omega)$ are the frequency-dependent dielectric
permittivities of gold and Si, and the following notations are
introduced
\begin{equation}
q_l=\left(\frac{\xi_l^2}{c^2}+k_{\bot}^2\right)^{1/2},
\quad
k_l^{(k)}=\left[\varepsilon^{(k)}(i\xi_l)
\frac{\xi_l^2}{c^2}+k_{\bot}^2\right]^{1/2}.
\label{eq10}
\end{equation}

The dielectric permittivities of gold and of high-resistivity Si
in the absence of laser light were computed \cite{13,46} by means
of the dispersion relation
\begin{equation}
\varepsilon^{(k)}(i\xi)=1+
\frac{2}{\pi}\int_{0}^{\infty}
d\omega
\frac{\omega\mbox{Im}\varepsilon^{(k)}(\omega)}{\omega^2+\xi^2},
\label{11}
\end{equation}
\noindent
where $\mbox{Im}\varepsilon^{(k)}(\omega)$ are taken from the tabulated
optical data for the complex index of refraction \cite{41}.
High-precision results for $\varepsilon^{(1)}(i\xi)$ (gold) are
presented in Ref.~\cite{46}. For high-resistivity Si the
behavior of $\varepsilon^{(2)}(i\xi)$ as a function of $\xi$ is
shown by the long-dashed line in Figs.~\ref{epsilon}a and
\ref{epsilon}b. In particular $\varepsilon^{(2)}(0)\approx 11.66$.

On irradiation of the Si membrane by light, the equilibrium value of the
carrier density is rapidly established during a period
of time much shorter than the duration of the laser pulse. Therefore,
we assume that there is an equilibrium concentration of pairs (electrons
and holes) when the light is incident. Thus, in the presence of laser
radiation, the dielectric permittivity of Si along the imaginary
frequency axis can be represented in the commonly used form
\cite{37,38,380,38b,40,41}
\begin{equation}
\varepsilon_l^{(2)}(i\xi)=\varepsilon^{(2)}(i\xi)
+\frac{\omega_p^{(e)}{\vphantom{\omega_p^{(e)}}}^2}{\xi\left[\xi
+\gamma^{(e)}\right]}
+\frac{\omega_p^{(p)}{\vphantom{\omega_p^{(e)}}}^2}{\xi\left[\xi
+\gamma^{(p)}\right]},
\label{eq12}
\end{equation}
\noindent
where $\omega_p^{(e,p)}$ and $\gamma^{(e,p)}$ are the plasma
frequencies and the relaxation parameters for electrons and holes,
respectively.

The values of the relaxation parameters
$\gamma^{(e)}\approx 1.8\times 10^{13}\,$rad/s and
$\gamma^{(p)}\approx 5.0\times 10^{12}\,$rad/s
can be found in Ref.~\cite{38b}. The plasma frequencies can be
calculated from
\begin{equation}
\omega_p^{(e,p)}=\left(
\frac{ne^2}{m_{e,p}^{\ast}\varepsilon_0}\right)^{1/2},
\label{eq13}
\end{equation}
\noindent
where the effective masses are \cite{38b} $m_p^{\ast}=0.2063m_e$,
$m_e^{\ast}=0.2588m_e$, $m_e$ is the electron mass, and $n$ is the
concentration of charge carriers.

The value of $n$ for the different
absorbed powers can be calculated in the following way. First, we
note that for a membrane of $d=4\,\mu$m thickness $n$ does not depend
on the depth.
The reason is that a uniform concentration in this direction is
established even more rapidly than the  equilibrium
discussed above \cite{38b}.
In fact the assumption on an uniform charge-carrier density
in the Si membrane is justified due to the long carrier diffusion
lengths and the ability to obtain almost defect free surfaces
in silicon through hydrogen passivation \cite{Yabl}.
Next, we approximately model the central part of the Gaussian beam
of diameter $w$
by a uniform cylindrical beam of the same diameter.
The power contained in this cylindrical beam, $P_w^{\rm eff}$,
is equal to the power in the central part of the Gaussian beam with
a diameter $w$. Elementary calculation using the Gaussian distribution
leads to $P_w^{\rm eff}=0.393P^{\rm eff}$.
The power $P_w^{\rm eff}$ is absorbed uniformly in the central
part of the Si membrane of diameter $w$ having a volume $V=\pi w^2d/4$.
Incidentally, the central region of the membrane with a diameter $w$
contributes almost 100\% (99.9999\% \cite{47}) of the total
Casimir force acting between a membrane and a sphere. At equilibrium, the
number of created charge carrier pairs per unit time per unit
volume $P_w^{\rm eff}/(\hbar\omega V)$, where
$\omega=3.66\times 10^{15}\,$rad/s is the frequency of Ar laser light,
is equal to the recombination rate of pairs per unit
volume $n/\tau$. Thus, at equilibrium
\begin{equation}
n=\frac{4P_w^{\rm eff}\tau}{\hbar\omega d\pi w^2}.
\label{eq14}
\end{equation}

Eqs.~(\ref{eq13}) and (\ref{eq14}) allow us to calculate the
densities of charge carriers
$n_{\rm a}=(2.1\pm0.4)\times 10^{19}\,\mbox{cm}^{-3}$,
$n_{\rm b}=(2.0\pm0.4)\times 10^{19}\,\mbox{cm}^{-3}$,
$n_{\rm c}=(1.4\pm0.3)\times 10^{19}\,\mbox{cm}^{-3}$
and the respective plasma frequencies
\begin{eqnarray}
&&
\omega_{p,{\rm a}}^{(e)}=(5.1\pm 0.5)\times 10^{14}\,\mbox{rad/s},
\quad
\omega_{p,{\rm a}}^{(p)}=(5.7\pm 0.6)\times 10^{14}\,\mbox{rad/s},
\nonumber \\
&&
\omega_{p,{\rm b}}^{(e)}=(5.0\pm 0.5)\times 10^{14}\,\mbox{rad/s},
\quad
\omega_{p,{\rm b}}^{(p)}=(5.6\pm 0.5)\times 10^{14}\,\mbox{rad/s},
\label{eq15} \\
&&
\omega_{p,{\rm c}}^{(e)}=(3.7\pm 0.4)\times 10^{14}\,\mbox{rad/s},
\quad
\omega_{p,{\rm c}}^{(p)}=(4.1\pm 0.4)\times 10^{14}\,\mbox{rad/s}
\nonumber
\end{eqnarray}
\noindent
in all measurements a, b, and c with different powers of the
absorbed laser light. In the calculations of charge carrier densities
using Eq.~(\ref{eq14}) we have used
$\tau_{\rm a}=\tau_{\rm b}=0.38\pm 0.03\,$ms
and $\tau_{\rm c}=0.47\pm 0.01\,$ms in accordance with the measurement
results in Sec.~IIIB,  taking into account the fact that $\tau$ decreases
when $n$ increases.
Recall that $\tau_{\rm a}$ and $\tau_{\rm b}$ were obtained from first
0.5\,ms of the time decay.
Our value for $\tau_{\rm c}$ obtained using the whole
5\,ms decay may lead to a minor underestimation of the carrier
density, a fact included in the resulting 21\% error
in the value of $n_{\rm c}$.
Note that the above values of the relaxation parameters $\gamma^{(e)}$
and $\gamma^{(p)}$ do not depend on the absorbed power \cite{38b} and
can be used in all measurements.

In Fig.~\ref{epsilon}a the dielectric permittivity of Si in the
presence of laser radiation (\ref{eq12}) is shown by solid lines a, b
and c as a function of imaginary frequency for the measurements with
different absorbed powers a, b and c, respectively.
The lines a and b in Fig.~\ref{epsilon}a almost coincide.
The region around the first Matsubara frequency $\xi_1$ at $T=300\,$K
is shown in Fig.~\ref{epsilon}b on an enlarged scale.

The obtained values of $\varepsilon^{(1)}(i\xi)$,
$\varepsilon^{(2)}(i\xi)$, and $\varepsilon_l^{(2)}(i\xi)$ were
substituted in the Lifshitz formula (\ref{eq8}) and the difference of
the Casimir forces $\Delta F_C(z)$ from Eq.~(\ref{eq5}) in the presence
and in the absence of laser light was computed at the laboratory
temperature $T=300\,$K. Note that there is discussion in the
literature on the correct value of the reflection coefficient for gold
$r_{\bot}^{(1)}(0,k_{\bot})$ at zero frequency (see, e.g.,
Refs.~\cite{12,13,13a,48,49,50,50aa}). Our calculation, however, does not
depend on chosen value of $r_{\bot}^{(1)}(0,k_{\bot})$ because
in Eq.~(\ref{eq8}) it is multiplied by $r_{\bot}^{(2)}(0,k_{\bot})=0$
for the silicon.
In the absence of light the latter equality holds for any true dielectric
with finite static dielectric permittivity. In the presence of light
the equality $r_{\bot,l}^{(2)}(0,k_{\bot})=0$ also holds true as is seen
from the substitution of Eq.~(\ref{eq12}) in Eq.~(\ref{eq9}).
In both cases at zero frequency only the transverse magnetic mode of
the electromagnetic field contributes to the result. Note that for Si
in the absence and in the presence of light for the transverse magnetic
mode
\begin{equation}
r_{\|}^{(2)}(0,k_{\bot})=\frac{\varepsilon^{(2)}(0)-1}{\varepsilon^{(2)}(0)+1}
\quad \mbox{and} \quad
r_{\|,l}^{(2)}(0,k_{\bot})=1,
\label{eq15a}
\end{equation}
\noindent
respectively.
Finally the Lifshitz formula (\ref{eq8}) was used to
compute the difference in the Casimir forces at all experimental
separations $z_i$ ($1\leq i\leq 1209$) and for the three measurements
performed at different absorbed powers.

The results of these calculations should be corrected for the presence
of surface roughness \cite{MM}.
The stochastic roughness on our test bodies can be
taken into account using the procedure presented in detail in
Refs.~\cite{13,18,37,38}. First, the zero roughness levels on both
gold ($H_0^{(1)}$) and Si ($H_0^{(2)}$) are determined from
\begin{equation}
\sum\limits_{k=1}^{33}\left(H_0^{(1)}-h_k\right)v_k=
\sum\limits_{l=1}^{17}\left(H_0^{(2)}-h_l\right)v_l=0,
\label{eq16}
\end{equation}
\noindent
where the heights $h_k,\,h_l$ and the fractions of the surfaces covered by
roughness with these heights are given in Tables~I and II,
respectively. From
Eq.~(\ref{eq16}) it follows $H_0^{(1)}=20.0\,$nm, $H_0^{(2)}=1.1\,$nm.
The absolute separation $z$ between the test bodies is in fact
measured between the zero roughness levels. Then the theoretical
values of the difference Casimir force with account of the surface
roughness are calculated as the geometric averaging
\begin{equation}
\Delta F_C^{\rm theor}(z_i)=\sum\limits_{k=1}^{33}\sum\limits_{l=1}^{17}
v_kv_l\Delta F_C(z_i+H_0^{(1)}+H_0^{(2)}-h_k-h_l),
\label{eq17}
\end{equation}
\noindent
where $\Delta F_C(z)$ was computed by the Lifshitz formula for perfectly
shaped bodies with and without light on a Si membrane. In the present
experiments the contribution from roughness correction is very small. Thus,
at $z=100\,$nm it contributes only 1.2\% of the calculated
$\Delta F_C^{\rm theor}(z)$. At $z=150\,$nm the contribution from
surface roughness decreases to only 0.5\% of the calculated force
difference. Similar to Refs.~\cite{13,18,38} it is easily seen that
the contribution from the nonadditive, diffraction-type effects to
roughness correction [which is not taken into account in Eq.~(\ref{eq17})]
is negligibly small.

The results of the numerical computations of the difference Casimir
force between rough surfaces $\Delta F_C^{\rm theor}(z)$ are shown as
solid lines in Fig.~\ref{result1},a-c for the measurements with different
powers of the absorbed laser light. They are in a very good agreement
with the experimental data shown by dots in the same figures (see the
following subsections for the quantitative measure of agreement between
experiment and theory).

For completeness, we present also the results of theoretical
computations using the Lifshitz formula at zero temperature. They are
obtained from Eq.~(\ref{eq8}) by changing the discrete Matsubara
frequencies $\xi_l$ for continuous $\xi$ and by replacement of
the summation for integration
\begin{equation}
k_BT\sum\limits_{l=0}^{\infty}\left(1-\frac{1}{2}\delta_{l0}\right)
\to\frac{\hbar}{2\pi}\int_{0}^{\infty}d\xi.
\label{eq18}
\end{equation}
\noindent
Following the same procedure as at $T=300\,$K, we first calculate
$\Delta F_C(z;T=0)$ using the Lifshitz formula and then
find $\Delta F_C^{\rm theor}(z;T=0)$ including the
effect of surface roughness with Eq.~(\ref{eq17}).
The results of these computations are shown as short-dashed
lines in Fig.~\ref{result1},a-c. As is seen from the figure, in all
cases the short-dashed lines describe a slightly larger magnitude of the
Casimir force difference than at $T=300\,$
in rather good agreement with
the experimental data shown as dots (see the next subsections for further
discussion).

\subsection{Analysis of theoretical errors}

The theoretical errors in the computation of the Casimir force acting
between a sphere and a membrane were discussed in detail in Refs.~\cite{18,38}.
The major source of the theoretical uncertainty in this
experiment is the
error in the concentration of charge carriers $n$ when the light is on.
From Sec.~IVA, this error is of about 20\%. Calculations using
the Lifshitz formula show that the resulting
relative error in the difference
Casimir force,
$\delta_1\left(\Delta F_C^{\rm theor}\right)\approx 0.12$,
i.e.,  is equal to approximately 12\%
and does not depend on separation. The error due to uncertainty
of experimental separations $z_i$, in which the theoretical values
$\Delta F_C^{\rm theor}$ should be computed, is equal to $3\Delta z/z$
and takes the maximum value of 3\% of the Casimir force at the shortest
separation of $z=100\,$nm (recall that according to Sec.~IIIA
$\Delta z=1\,$nm). This leads to only 2\% error in the difference of the
Casimir force at $z=100\,$nm
[so that $\delta_2\left(\Delta F_C^{\rm theor}\right)\approx 0.02$]
and to smaller errors at larger separations.
The other sources of theoretical errors, discussed in Refs.~\cite{18,38},
like sample-to-sample variation of the tabulated optical data in Au,
use of the proximity force theorem, patch potentials, nonlocal effects
and finite thickness of the gold coating on the sphere contribute
negligible amounts to the error in  $\Delta F_C^{\rm theor}$.
Thus, for example, using the Lifshitz formula for a polystyrene sphere
covered by a gold layer of 82\,nm thickness instead of Eq.~(\ref{eq8})
written for a solid gold sphere, we would get only a 0.03\% decrease
in the Casimir force magnitude.

A specific new uncertainty which is present in this
experiment is connected with the pressure of light transmitted through the
membrane and incident on the bottom of the
sphere (see Sec.~II). This effect is present
only during the light phase of the pulse train and can be easily
estimated. The maximum intensity of the laser light incident on a sphere
section
with radius $0\leq r\leq R$ parallel to the membrane is
\begin{equation}
I(r)=\frac{2\alpha P^{\rm eff}}{\pi w^2}e^{-\frac{2r^2}{w^2}},
\label{eq19}
\end{equation}
\noindent
where $\alpha$ is the fraction of the absorbed power transmitted
through the membrane.
The value of $\alpha$ is given by
\begin{equation}
\alpha=re^{-d/l_{\rm opt}}\approx 0.00641,
\label{eq20}
\end{equation}
\noindent
where $l_{\rm opt}=1\,\mu$m (see Sec.~II) and the transmission coefficient
$r\approx 0.35$.

The force due to light pressure acting on the sphere in spherical
coordinates takes the form
\begin{equation}
F_p=\frac{4\pi R^2}{c}\int_{0}^{\pi/2}d\vartheta
I(R\sin\vartheta)\cos^2\vartheta\sin\vartheta.
\label{eq21}
\end{equation}
\noindent
Substituting Eq.~(\ref{eq19}) in Eq.~(\ref{eq21}) and integrating, one
obtains
\begin{equation}
F_p=\frac{2\alpha P^{\rm eff}}{c}\left[1-e^{-\frac{2R^2}{w^2}}
\frac{\sqrt{\pi}w\mbox{Erfi}(\sqrt{2}R/w)}{2\sqrt{2}R}\right],
\label{eq22}
\end{equation}
\noindent
where Erfi$(z)$ is the imaginary error function.

For the absorbed powers used in three experiments ($P^{\rm eff}=9.3$, 8.5,
and 4.7\,mW, respectively), Eq.~(\ref{eq22}) leads to the following
maximum forces which may act on the sphere due to light pressure:
$F_p=0.085$, 0.078 and 0.043\,pN.
The force due to light pressure can be taken into account as one
more error in the theoretical evaluation of the Casimir force
difference $\Delta F_C^{\rm theor}$.
At a separation $z=100\,$nm the respective relative error,
$\delta_3\left(\Delta F_C^{\rm theor}\right)$, is equal to
2.3, 2.7, and 1.5\%  for the three absorbed powers.
At $z=200\,$nm the relative theoretical error in $\Delta F_C^{\rm theor}$
due to
light pressure increases up to 8.9, 8.7, and 5.0\%, respectively.

All three errors discussed above can be considered as the random
quantities described by the same distribution law which is close to
a uniform distribution. For this reason the statistical criterion
\cite{45} used in Sec.~IIIC can be applied once more, giving
the total relative theoretical error in the difference
Casimir force
\begin{equation}
\delta^{\rm tot}(\Delta F_C^{\rm theor})=\min\left\{
\sum\limits_{i=1}^{q}\delta_i(\Delta F_C^{\rm theor}),
k_{\beta}^{(q)}\sqrt{
\sum\limits_{i=1}^{q}\left[
\delta_i(\Delta F_C^{\rm theor})\right]^2}\right\}
\label{eq22a}
\end{equation}
\noindent
with  $q=3$ and
$k_{0.95}^{(3)}=1.12$.
The resulting total absolute theoretical error,
\begin{equation}
\Delta^{\!\rm tot}\left(\Delta F_C^{\rm theor}\right)=
|\Delta F_C^{\rm theor}|
\delta^{\rm tot}\left(\Delta F_C^{\rm theor}\right),
\label{eq22aa}
\end{equation}
\noindent
is presented in Fig.~\ref{erTh} as a
function of separation for the three experiments with decreasing power
of the absorbed laser light (lines a, b, and c, respectively). As is
seen from this figure, the total theoretical errors for the
measurements a and b are almost equal, and for the measurement
c this error is slightly lower. The relative total theoretical error
changes from 13.5 to 13.7\% at $z=100\,$nm and from 13.7 to 14.4\%
at $z=140\,$nm for the three different absorbed powers.
At $z=200\,$nm the relative total theoretical error ranges from
14.9 to 17.2\% for the different absorbed powers.

\subsection{Measure of agreement between experiment and theory}

In the foregoing we have independently found the total experimental
(Sec.~IIIC) and theoretical (Sec.~IVB) errors in the difference of the
Casimir force in the presence and in the absence of laser light
excited carriers at
95\% confidence. To compare experiment with theory, we consider the
quantity $\Delta F_C^{\rm theor}-\Delta F_C^{\rm expt}$
and determine its absolute error $\Xi_{0.95}(z)$ as a function of
separation at the confidence of 95\%. This can be done in the same
procedure as in Refs.~\cite{13,38,45a} applying the statistical criterion
\cite{45} and using the data in Figs.~\ref{random} and \ref{erTh}
\begin{equation}
\Xi_{\beta}=\min\left\{
\Delta^{\rm tot}(\Delta F_C^{\rm expt})+
\Delta^{\rm tot}(\Delta F_C^{\rm theor}),
k_{\beta}^{(2)}\sqrt{
\left[\Delta^{\!\rm tot}(\Delta F_C^{\rm expt})\right]^2+
\left[\Delta^{\!\rm tot}(\Delta F_C^{\rm theor})\right]^2
}\right\}.
\label{eq22b}
\end{equation}
\noindent
Here $k_{0.95}^{(2)}=1.10$.
The resulting confidence intervals $[-\Xi_{0.95}(z),\Xi_{0.95}(z)]$
are shown in Fig.~\ref{result2},a-c as the solid lines for the three
measurements with the largest, intermediate, and smallest powers,
respectively.

The differences between the theoretical values of
$\Delta F_C^{\rm theor}$ (computed in Sec.~IVA at $T=300\,$K) and
experimentally measured $\langle\Delta F_C^{\rm expt}\rangle$ are
shown in Fig.~\ref{result2} by dots labeled 1 (once again dots in
Fig.~\ref{result2},a-c are related to the three measurements with
different power). As is seen from Fig.~\ref{result2}, practically all
dots labeled 1 are well inside the confidence intervals at all
separation distances. This means that the Lifshitz theory at nonzero
temperature, using the dielectric permittivity of high-resistivity
Si $\varepsilon^{(2)}(i\xi)$ in the absence of laser light and the
dielectric permittivity $\varepsilon_l^{(2)}(i\xi)$ given by
Eq.~(\ref{eq12}) in the presence of light, is consistent with
experiment. The consistency of the experiment with the theory is
preserved when the theoretical values of $\Delta F_C^{\rm theor}$
are computed at zero temperature (see the short-dashed lines in
Fig.~\ref{result1},a-c and the discussion in Sec.~IVA). The reason is
that the thermal correction to the Casimir force in the region of
small separations under consideration is practically negligible and
the thermal effect cannot be resolved taking into consideration
the experimental and theoretical errors reported above.

For illustrative purposes, the agreement between experiment and
theory is presented in a more standard form in Fig.~\ref{result3}.
Here a more narrow separation interval from 100 to 150\,nm is
considered and each third experimental point from the measurement b is
plotted together with its error bars
$\left[\pm\Delta z,\pm\Delta^{\rm \! tot}
\left(\Delta F_C^{\rm expt}\right)\right]$
shown as crosses (there are too many points to present all of them in
this form). The theoretical force difference $\Delta F_C^{\rm theor}$
computed by the Lifshitz formula at $T=300\,$K is shown by the solid
line. It is seen that the experimental data are in a very good agreement
with the theory in confirmation of the conclusion made above using
Fig.~\ref{result2}.

\section{Problem of  dc conductivity of high-resistivity Si in
the Lifshitz theory}

In Sec.~IVA the dielectric response of high-resistivity Si in the
absence of excitation laser light was described by the function
$\varepsilon^{(2)}(i\xi)$ having a finite static value
$\varepsilon^{(2)}(0)\approx 11.66$. It is common knowledge, however, that
dielectrics have some nonzero dc conductivity $\sigma_0$ at any nonzero
temperature. This conductivity decreases with the decrease of temperature
as $\sigma_0\sim\exp(-b/T)$, where $b$ can be expressed in terms of the
band gap or dopant activation
energy. To take the dc conductivity into account
in the Lifshitz theory, the dielectric
permittivity of Si along the imaginary frequency axis
$\varepsilon^{(2)}(i\xi)$ used in Sec.~IVA should be replaced with
\begin{equation}
{\tilde{\varepsilon}}^{(2)}(i\xi)=\varepsilon^{(2)}(i\xi)+
\frac{{\tilde{\omega}}_p^{(p)}{\vphantom{{\tilde{\omega}}_p^{(p)}}}^2}{\xi
\left[\xi+\gamma^{(p)}\right]}.
\label{eq23}
\end{equation}
\noindent
The value of the plasma frequency in Eq.~(\ref{eq23}) is found
by substituting the concentration of carrier density
$\tilde{n}\approx 5\times 10^{14}\,\mbox{cm}^{-3}$ (see Sec.~II)
in Eq.~(\ref{eq13}) with the result
${\tilde{\omega}}_p^{(p)}\approx 2.8\times 10^{12}\,$rad/s.
Note that for $n\leq 1.0\times 10^{17}\,\mbox{cm}^{-3}$ the value of the
relaxation parameter has an insignificant effect on the magnitude of the
Casimir force \cite{38b}. Because of this in Eq.~(\ref{eq23}) the same
value of $\gamma^{(p)}$ as in Eq.~(\ref{eq12}) is used.
The behavior
of ${\tilde{\varepsilon}}^{(2)}$ as a function of $\xi$ is plotted in
Fig.~\ref{epsilon}a by the short-dashed line.

The presence of some low dc conductivity in dielectric materials
was used in Refs.~\cite{6,51} to obtain a large effect of the
van der Waals friction which could bring the observations of Ref.~\cite{5}
in agreement with theory.
In Ref.~\cite{7} for two dielectric plates and in \cite{IJMPA,48a} for
one metal and one dielectric plate
it was proved, however,  that the inclusion of the dc
conductivity for dielectrics into the Lifshitz theory leads to the
violation of the third law of thermodynamics (the Nernst heat theorem).
Thus, it is not acceptable from a theoretical point of view.

Our experiments on the modification of the Casimir force with laser
pulses clarify the problem whether or not the
dc conductivity of high-resistivity Si should be
taken into account in the Lifshitz theory of the Casimir and van der
Waals forces. For this purpose, we have completely repeated the
theoretical computations of the difference Casimir force made in Sec.~IVA
replacing the dielectric permittivity of Si
${{\varepsilon}}^{(2)}(i\xi)$, used
there, for ${\tilde{\varepsilon}}^{(2)}(i\xi)$ given in Eq.~(\ref{eq23}).
The obtained theoretical results for
$\Delta{\tilde{F}}_C^{\rm theor}$
versus separation are shown by the long-dashed lines
in Fig.~\ref{result1},a-c for all the three measurements with different
powers of the absorbed light. As is seen in Fig.~\ref{result1}, all
the long-dashed lines are far outside
both the experimental data shown as dots and from
the solid lines calculated using the Lifshitz theory
disregarding  dc conductivity of high-resistivity Si
at the laboratory temperature. Notice that the computational results
at $T=0$ (shown by the short-dashed lines in Fig.~\ref{result1})
do not depend on whether the dc conductivity is included in
the dielectric permittivity used to describe the
high-resistivity Si.

To make a quantitative conclusion on the measure of agreement between
the data and two models with and without inclusion
of dc conductivity of
high-resistivity Si,
we have plotted in Fig.~\ref{result2},a-c
the differences
$\Delta{\tilde{F}}_C^{\rm theor}-\langle\Delta F_C^{\rm expt}\rangle$,
where $\Delta{\tilde{F}}_C^{\rm theor}$ was computed including the dc
conductivity according to Eq.~(\ref{eq23}). These differences are shown
as dots labeled 2 in Fig.~\ref{result2},a-c. As is seen in
Fig.~\ref{result2},a,b, the model with included dc conductivity of
high-resistivity Si is excluded experimentally at 95\%
confidence within the region from 100 to 250\,nm. From Fig.~\ref{result2},c
it follows that this model is excluded at 95\% confidence within
the separations region from 100 to 200\,nm.

The same conclusion, that the model of high-resistivity Si,
 which includes dc conductivity, is inconsistent with our
experiments on the optically modulated
Casimir force, is confirmed also in Fig.~\ref{result3},
where the  quantity $\Delta{\tilde{F}}_C^{\rm theor}$ versus separation
is plotted as the dashed line. It can be clearly observed that the dashed line is
not only far away from the solid line based on theory neglecting the Si
dc conductivity in the absence of excitation light, but is also distant
from all error bars representing the experimental data.

The physical explanation for
 the deviations of the long-dashed lines from the solid lines in
Figs.~5,a--c and 10 is as follows. When the dc conductivity
of Si is taken into account, the equalities
$r_{\bot}^{(2)}(0,k_{\bot})=r_{\bot,l}^{(2)}(0,k_{\bot})=0$ follow from the
substitution of Eqs.~(\ref{eq12}) and (\ref{eq23}) in Eq.~(\ref{eq9}).
Once again, at zero frequency only the transverse magnetic mode contributes
to the result. Here, however, for Si both in the absence and in the
presence of light the equations
\begin{equation}
{\tilde{r}}_{\|}^{(2)}(0,k_{\bot})=1
\quad \mbox{and} \quad
{\tilde{r}}_{\|,l}^{(2)}(0,k_{\bot})=1
\label{eq29}
\end{equation}
\noindent
hold. It is exactly this change in the magnitude of the transverse magnetic
reflection coefficient $r_{\|}^{(2)}(0,k_{\bot})$, as given in Eq.~(\ref{eq15a}),
with ${\tilde{r}}_{\|}^{(2)}(0,k_{\bot})$ in Eq.~(\ref{eq29}) leads to the deviation of
the long-dashed lines from the respective solid lines in Figs.~5,a--c and 10.
It seems somewhat surprising that the use of the permittivity
${\tilde{\varepsilon}}^{(2)}(i\xi)$ in Eq.~(\ref{eq23}), which can be considered
as a more exact than $\varepsilon^{(2)}(i\xi)$, leads to the discrepancy between
experiment and theory. This is, in fact, one more observation that there are
puzzles concerning the applicability of the Lifshitz theory to real materials.
In the case of metals, the Drude description of conduction electrons in the
thermal Casimir force was excluded experimentally in the series of
experiments \cite{12,13,13a}. It also leads to the contradiction with
the Nernst heat theorem for perfect crystal lattices \cite{50}.
For metals, the deviation of the experimental results from the Drude
model approach and the violation of the Nernst theorem are explained
by the vanishing contribution from the transverse electric mode
at
zero frequency. The present experiment dealing with semiconductors is
not sensitive enough to detect this effect. The effect reported here is
novel and arises due to the difference in the contributions
of the zero-frequency transverse magnetic mode. These contributions,
as was shown above, depend on whether or not the dc conductivity
of Si in the absence of light is taken into account.

\section{Conclusions and discussion}

In this paper we demonstrate that it is possible to control the Casimir
force between the gold coated sphere and Si membrane by the irradiation
of Si with laser pulses.
On absorption of light, the carrier density increases leading to an
increase in the magnitude of the Casimir force.
This change in the
Casimir forces was investigated as a function of separation between the
test bodies and the power of the absorbed light.
The experiments were performed with a specially prepared single crystal
Si membrane in an oil-free vacuum chamber using an AFM.
The developed calibration procedure
permitted measurement of the difference Casimir force of
the order of 1\,pN with
a relative experimental error at the shortest separation of 100\,nm
varying from 10 to 20\% for the measurements performed at different absorbed
powers. At a separation of 180\,nm the relative experimental error in
different measurements varies from 25 to 33\%. All errors were determined
at 95\% confidence. The obtained experimental results demonstrate
the ability  to modulate
the van der Waals and Casimir forces in micro- and nanoelectromechanical
devices by irradiation with laser light. These are
pioneering experiments where the modification of the Casimir force acting
between the test bodies was achieved due to the influence of some
external factor other than the change of separation distance.

The experimental results were compared with the results of theoretical
computations using the Lifshitz theory at both zero and
nonzero temperature.
The Si membrane in the absence of laser light had a carrier density of
approximately $5\times 10^{14}\,\mbox{cm}^{-3}$.
In the first model, the
dielectric permittivity of high-resistivity Si
was described with a finite static value.
In the presence of laser light the Si had charge carriers pair
densities varying from $2.1\times 10^{19}\,\mbox{cm}^{-3}$ to
$1.4\times 10^{19}\,\mbox{cm}^{-3}$ depending on the radiation power
absorbed by the sample and was described by the permittivity
in Eq.~({\ref{eq12}).
The total theoretical error varied from 13.5 to 13.7\% at $z=100\,$nm
and from 14.9 to 17.2\% at $z=200\,$nm depending on the absorbed power.
The main contribution to this error was given by the uncertainty in the number
of charge carriers in the presence of laser light. The experimental and
theoretical results were found to be consistent over the whole
measurement range taking into account the experimental and
theoretical errors both at laboratory temperature $T=300\,$K and
at zero temperature.

The same experimental data were compared with the Lifshitz theory
using a second model of high-resistivity Si which
includes the dc conductivity of the Si membrane
in the absence of laser radiation.
In this case the dielectric permittivity of Si in the absence of radiation
is represented by Eq.~(\ref{eq23}) and goes to infinity when the frequency
goes to zero. The detailed comparison leads to
the conclusion that this model
is excluded by the experiment at 95\% confidence
if computations are performed at the laboratory temperature
$T=300\,$K.
The difference in the force magnitudes when conductivity at zero frequency
is absent or present arises from different contributions of the
transverse magnetic modes of the electromagnetic field reflected from the
Si surface.
The physical
explanation of our results can be understood in Fig.~\ref{epsilon}a.
As is seen from this figure,
 the short-dashed line representing the dielectric permittivity
of high-resistivity Si with included dc conductivity is located
far to the left of the first Matsubara frequency $\xi_1$ and does not
belong to the region of frequencies contributing to the force.
At the same time, the Lifshitz theory at zero temperature using the
model of high-resistivity Si with included dc conductivity
remains experimentally consistent.

Thus, we can infer that the Lifshitz theory at nonzero
temperature using the model of high-resistivity semiconductors
and dielectrics with included conductivity properties at
zero-frequency is inconsistent with our experiments.
It is notable that just this theoretical approach was demonstrated
\cite{7,IJMPA,48a,JPA} to lead to the violation of the third law of
thermodynamics (the Nernst heat theorem). To avoid contradictions
with thermodynamics and experiment one should follow the
originators of the Lifshitz theory \cite{23,24} who described
dielectrics by a model with a finite static dielectric
permittivity in computations of the van der Waals and Casimir
forces at nonzero temperature (the same model was used in the recent paper
\cite{64a} on the thermal effect in the Casimir-Polder force).
This suggests that the theory of van der Waals and Casimir forces between real
materials requires further investigation.
Although we are still lacking
a fundamental explanation of why the Lifshitz theory does
not admit inclusion of the conductivity properties of
high-resistivity materials at zero frequency, this
prescription on how to perform computations in an
experimentally and thermodynamically consistent way is
topical for numerous applications of the van der Waals
and Casimir forces ranging from condensed matter physics
and nanotechnology to the theory of fundamental interactions.
The experimentally demonstrated phenomenon of modulation
of the Casimir force through optical modification of
charge-carrier density will be used in the design and
function of micro- and nanoelectromechanical devices such
as nanoscale actuators, micromirrors and nanotweezers.

%%%%%%%%%%%%%%%%%%%%%%%%%%%%%%%%%%%%%%%%%
\section*{Acknowledgment}
G.L.K. and V.M.M. are grateful to the Department of Physics
and Astronomy of
the University of California (Riverside) for its kind hospitality.
The instrumentation in this work was supported by the NSF
Grant PHY0653657. Theoretical calculations and personnel were
supported by the
DOE grant DE-FG02-04ER46131.

%%%%%%%%%%%%%%%%%%%%%%%%%%%%%%%%%%%%%%%%%%

%%%%%%%%%%%%%%%%%%%%%%%%%%%%%%%%%%%%%%%%%
\begingroup
\squeezetable
\begin{table}
\caption{Fractions $v_{k}$ of Au surface  covered by
roughness with heights $h_{k}$.}
\begin{ruledtabular}
\begin{tabular}{lcc}
$k$ & $h_k\,$(nm) & $v_k$ \\
\hline
1&0&$7\times 10^{-5}$ \\
2&1&$6.0\times 10^{-4}$ \\
3&2&$6.3\times 10^{-4}$ \\
4&3&$7.0\times 10^{-4}$ \\
5&4&$5.0\times 10^{-4}$ \\
6&5&$2.1\times 10^{-3}$ \\
7&6&$1.4\times 10^{-3}$ \\
8&7&$4.0\times 10^{-3}$ \\
9&8&$7.0\times 10^{-3}$ \\
10&9&$8.0\times 10^{-3}$ \\
11&10&$1.2\times 10^{-2}$ \\
12&11&$1.3\times 10^{-2}$\\
13&12&$1.3\times 10^{-2}$\\
14&13&$2.0\times 10^{-2}$\\
15&14&$2.7\times 10^{-2}$\\
16&15&$3.6\times 10^{-2}$ \\
17&16&$4.4\times 10^{-2}$ \\
18&17&$6.0\times 10^{-2}$ \\
19&18&$7.4\times 10^{-2}$\\
20&19&$8.6\times 10^{-2}$\\
21&20&$8.7\times 10^{-2}$\\
22&21&$8.8\times 10^{-2}$\\
23&22& 0.111 \\
24&23& 0.1 \\
25&24&$7.7\times 10^{-2}$\\
26&25&$5.4\times 10^{-2}$\\
27&26&$3.5\times 10^{-2}$ \\
28&27&$2.0\times 10^{-2}$ \\
29&28&$9.0\times 10^{-3}$\\
30&29&$4.0\times 10^{-3}$\\
31&30&$3.0\times 10^{-3}$\\
32&31&$1.0\times 10^{-3}$\\
33&32&$1.0\times 10^{-3}$
\end{tabular}
\end{ruledtabular}
\end{table}
\endgroup
%%%%%%%%%%%%%%%%%%%%%%%%%%%%%%%%%%%%%%%
\begingroup
\squeezetable
\begin{table}
\caption{Fractions $v_{l}$ of Si surface covered by
roughness with heights $h_{l}$.}
\begin{ruledtabular}
\begin{tabular}{lcc}
$l$ &$h_l\,$(nm)& $v_l$ \\
\hline
1&0&$5.0\times 10^{-4}$ \\
2&0.18&$1.5\times 10^{-3}$ \\
3&0.28&$2.0\times 10^{-3}$ \\
4&0.38&$4.0\times 10^{-3}$ \\
5&0.48&$7.0\times 10^{-3}$\\
6&0.58&$5.0\times 10^{-3}$ \\
7&0.68&$1.0\times 10^{-2}$ \\
8&0.78&$4.0\times 10^{-2}$ \\
9&0.88&$8.0\times 10^{-2}$ \\
10&0.98&0.15 \\
11&1.08&0.22 \\
12&1.18&0.215 \\
13&1.28&0.147 \\
14&1.38&$8.3\times 10^{-2}$ \\
15&1.48&$2.4\times 10^{-2}$ \\
16&1.58&$9.0\times 10^{-3}$ \\
17&1.68&$2.0\times 10^{-3}$
\end{tabular}
\end{ruledtabular}
\end{table}
\endgroup
%%%%%%%%%%%%%%%%%%%%%%%%%
%%%%%%%%%%%%%%%%%%%%%%%%%%%%%%%%%%
%%%__FIGURES__%%%%%%%%%%
%%%%%%%%%%%%%%
\begin{figure}
%\vspace*{0cm}
%\centerline{\includegraphics{figIr-1.ps}}
%\vspace*{-12cm}
\caption{\label{setup}
Schematic of the experimental setup, showing its main components
(see text).}
\end{figure}
%%%
\begin{figure}
%\centerline{\includegraphics{figIr-2.ps}}
%\vspace*{-8cm}
\caption{Fabrication process of Si membrane.
{\bf a}, The  Si substrate (coloured black) with a buried
SiO${}_2$ layer (white). {\bf b}, The substrate is mechanically
polished and oxidized, and
{\bf c}, a window in SiO${}_2$ is etched with
HF. {\bf d}, Next, TMAH is used to etch the Si. {\bf e},
Finally SiO${}_2$ layer is etched away in HF solution to
form a clean Si surface.
}
\end{figure}
%%%%
\begin{figure}
%\vspace*{-1cm}
%\centerline{\includegraphics{figIr-3.ps}}
%\vspace*{-12cm}
\caption{\label{deflSignal}
The deflection signal of the cantilever in response to the
dc voltage and square voltage pulse applied to the Si membrane
as a function of separation.}
\end{figure}
%%%%
\begin{figure}
%\vspace*{-1cm}
%\centerline{\includegraphics{figIr-4.ps}}
%\vspace*{-12cm}
\caption{\label{lifetime}
The change of the reflectivity after the termination of
the laser pulse.}
\end{figure}
%%%%
\begin{figure}
%\vspace*{1cm}
%\centerline{\includegraphics{figIr-5.ps}}
%\vspace*{-13cm}
\caption{\label{result1}
The differences of the Casimir forces in the presence and in the absence of
light versus separation for different absorbed powers:
(a) 9.3\,mW; (b) 8.5\,mW; (c) 4.7\,mW.
The measured differences $\langle\Delta F_C^{\rm expt}\rangle$ are shown
as dots, differences calculated using the Lifshitz formula at $T=300\,$K,
$\Delta F_C^{\rm theor}$, and at $T=0$, ${{\Delta}} F_C^{\rm theor}(T=0)$,
as the solid and short-dashed lines, respectively, and that
calculated including the
dc conductivity of high-resistivity Si,
 ${\Delta}{\tilde{F}}_C^{\rm theor}$,
as the long-dashed lines.}
\end{figure}
%%%%
\begin{figure}
%\vspace*{-6cm}
%\centerline{\includegraphics{figIr-6.ps}}
%\vspace*{-9cm}
\caption{\label{random}
The random errors (which are equal to the total) versus separation for
the measurements with the different  absorbed powers:
(a) 9.3\,mW; (b) 8.5\,mW; (c) 4.7\,mW.}
\end{figure}
%%%%
\begin{figure}
%\vspace*{-1cm}
%\centerline{\includegraphics{figIr-7.ps}}
%\vspace*{-10cm}
\caption{\label{epsilon}
(a) The dielectric permittivity of the Si membrane along
the imaginary frequency
axis in the absence of light (the long-dashed line is for the model of Si
with a finite
static permittivity and the short-dashed line includes dc
conductivity of high-resistivity Si) and in the presence of light
for different  absorbed powers [solid lines:
(a) 9.3\,mW; (b) 8.5\,mW; (c) 4.7\,mW].
(b) The same is shown on an enlarged scale in the region of the first
Matsubara frequency $\xi_1$.}
\end{figure}
%%%%
\begin{figure}
%\vspace*{-7cm}
%\centerline{\includegraphics{figIr-8.ps}}
%\vspace*{-9cm}
\caption{\label{erTh}
The total theoretical errors versus separation  for
measurements with different  absorbed powers:
(a) 9.3\,mW; (b) 8.5\,mW; (c) 4.7\,mW. }
\end{figure}
%%%%
\begin{figure}
%\vspace*{1cm}
%\centerline{\includegraphics{figIr-9.ps}}
%\vspace*{-13cm}
\caption{\label{result2}
Theoretical minus experimental differences in the Casimir force versus
separation  for
the measurements with different  absorbed powers:
(a) 9.3\,mW; (b) 8.5\,mW; (c) 4.7\,mW  are shown as dots. The results
computed at $T=300\,$K using the model
with a finite static permittivity of high-resistivity Si
are labeled 1 and that including the dc conductivity are labeled 2.
Solid lines show the 95\% confidence intervals.}
\end{figure}
%%%%
\begin{figure}
%\vspace*{-7cm}
%\centerline{\includegraphics{figIr-10.ps}}
%\vspace*{-9cm}
\caption{\label{result3}
The experimental differences in the Casimir force with their
experimental errors are shown as crosses. Solid and dashed lines
represent the theoretical differences computed
at $T=300\,$K using the model with a finite static
permittivity of high-resistivity Si and
that including the dc conductivity,
respectively.}
\end{figure}
%%%%

%%%%%%%%%%%%%%%%%%%%%%%%%%%%%%%%%
\end{document}